\documentstyle[12pt,aps,prc,epsfig,preprint]{revtex}
\tightenlines
\begin{document}
\begin{titlepage}

\title{Experiments with $\Xi^-$ atoms}
\author{C.J. Batty$^1$, E. Friedman$^2$, and A. Gal$^2$ \\
$^1$Rutherford Appleton Laboratory, Chilton, Didcot, Oxon, OX11 0QX, UK \\
$^2$Racah Institute of Physics, The Hebrew University, Jerusalem 91904,
Israel}
\maketitle

\begin{abstract}
Experiments with $\Xi^-$ atoms are proposed in order to study
the nuclear interaction of $\Xi$ hyperons. The production of
$\Xi^-$ in the ($K^{-}$, $K^{+}$) reaction, the $\Xi^-$ stopping in
matter, and its atomic cascade are incorporated within a realistic
evaluation of the results expected for $\Xi^-$ X-ray spectra across
the periodic table, using an assumed $\Xi$-nucleus optical potential
$V_{opt}$. Several optimal targets for  measuring the
strong-interaction shift and width of the X-ray transition to the
`last' atomic level observed are singled out: F, Cl, I, Pb. The
sensitivity of these observables to the parameters of $V_{opt}$ is
considered. The relevance of such experiments is discussed
in the context of strangeness $-2$ nuclear physics and multistrange
nuclear matter. Finally, with particular reference to searches for
the $H$ dibaryon, the properties of $\Xi^-d$ atoms are also discussed.
The role of Stark mixing, its effect on $S$ and
$P$ state capture of $\Xi^-$ by the deuteron, together with estimates
of the resulting probability for producing the $H$ dibaryon
are considered in detail.
\newline
$PACS$: 15.20.Jn, 36.10.Gv
\end{abstract}
\centerline{\today}
\end{titlepage}

\section{Introduction and Background}
\label{sec:int}

Very little is established experimentally or phenomenologically on
the interaction of $\Xi$ hyperons with nuclei.
Dover and Gal \cite{DGa83}, analyzing old emulsion data which had
been interpreted as due to $\Xi^-$ hypernuclei, obtained an attractive
$\Xi$-nucleus interaction with a nuclear potential well depth of
$V_{0}^{(\Xi)} = 21 - 24$ MeV. This range of values agrees well with
the theoretical prediction by the same authors \cite{DGa84} for $\Xi$
in nuclear matter, using model D of the Nijmegen group \cite{NRD77} to
describe baryon-baryon interactions in an SU(3) picture, in contrast
with the $\Xi$-nucleus repulsion obtained \cite{DGa84} using model F
\cite{NRD79}. Similar predictions were subsequently made with more
detailed $G$ matrix evaluations by Yamamoto {\it et al.}
\cite{YMH94,Yam95} who argued for a considerable $A$
dependence of $V_{0}^{(\Xi)}$, such that the well depth for light and
medium weight nuclei is significantly lower than for heavy nuclei
where it approaches the value calculated for nuclear matter. It should
be noted, however, that the predictions of the Nijmegen model D for
$V_{0}^{(\Xi)}$ are extremely sensitive to the value assumed for the
hard-core radius. Nevertheless, the confidence in the predictive power
of model D for this sector of strangeness  $-$2 hypernuclear
physics, at least qualitatively, is to a large extent due to its
success in yielding the attractive $\Lambda \Lambda$ interaction
necessary to reproduce the (so far) three known $\Lambda \Lambda$
binding energies (see Ref. \cite{YMH94} for a review of these
calculations).

If the interaction of $\Xi$ hyperons with nuclei is sufficiently
attractive to cause binding, as has been repeatedly
argued since the original work of Dover and Gal \cite{DGa83}, then
a rich source of spectroscopic information becomes available
and the properties of the in-medium $\Xi N$ interaction
can be extracted.  Bound states of $\Xi$ hypernuclei
would also be useful as a gateway to form double
$\Lambda$ hypernuclei \cite{MDG94,IFM94}. Finally, a minimum strength
for $V_{0}^{(\Xi)}$ of about 15 MeV is required to realize
the exciting possibility
of strange hadronic matter \cite{SDG93}; where protons, neutrons,
$\Lambda$s and $\Xi$s are held together to form a system which is
stable against strong-interaction decay.

Some new information on the $\Xi^{-}$ nucleus interaction has
been recently reported from ($K^-$, $K^+$) counter experiments at
the KEK proton synchrotron.
Fukuda {\it et al.} \cite{Fuk98} have shown fits to the very low energy part
(including the bound state region) of the $\Xi^{-}$ hypernuclear
spectrum in the $^{12}$C$(K^-, K^+)X$ reaction on a scintillating fiber
active target (experiment E224),  resulting in an
estimate of $V_{0}^{(\Xi)}$  between 15 to 20 MeV.
The experimental energy resolution of about 10 MeV in this experiment
was too poor to allow identification of any bound state peak structure
which could have given more definitive information on the well depth.
A somewhat cleaner and better resolved spectrum has been recently shown
\cite{May98} from the Brookhaven AGS experiment E885, but no analysis of
these data has yet been reported. An earlier KEK experiment (E176)
gave evidence for three events of stopped $\Xi^{-}$ in light emulsion
nuclei, each showing a decay into a pair of single $\Lambda$
hypernuclei. The first two events \cite{ABC93,ABC95} are consistent
energetically with a $\Xi^{-}$ {\it atomic} state in $^{12}$C bound by
$B_{\Xi^{-}}(^{12}{\rm C}) = 0.58 \pm 0.14~~{\rm MeV}$. However, this
value could only be ascribed to capture from the $1S$ state which is
estimated to occur in less than 1\% of the total number of captures.
This binding energy is distinctly larger than the calculated value
$B_{\Xi^-}^{2P}(^{12}{\rm C}){_{_{_{_<}}}\atop ^{^\sim }} 0.32$ MeV
for the 2$P$ state,
for a wide range of strong-interaction potentials. Moreover, the $\Xi^-$
capture probability in $^{12}$C from $P$ states is a few percent
at most. The most likely capture in $^{12}$C, as  discussed in
Sec. \ref{subsec:results}, occurs from atomic $D$ states.
The calculated binding energies of the atomic $3D$ states
for C, N, O emulsion nuclei are given in Table \ref{tab:Blight}
where it is seen that binding is essentially by the Coulomb potential.
The two examples for binding in the presence of a strong $\Xi$-nucleus
potential are for the $t\rho$ potential used in Sec. \ref{sec:exp}
and \ref{sec:tgts}
with the parameter $b_0=0.25+i0.04$ fm (potential 1)
and $b_0=0.19+i0.04$ fm (potential 2), corresponding to
$V_{0}^{(\Xi)} = $20.5 and 15.6 MeV, respectively, in $^{12}$C.
(The value used for Im$b_0$ is discussed in Sec. \ref{subsec:potls}
and \ref{sec:DD}). It is seen that the sensitivity of the binding
energies and strong interaction widths to the
$\Xi$-nucleus strong interaction is, for these examples,
of the order of 100 eV,
substantially smaller than a typical error of 100 keV incurred in
emulsion work. We point out that there exist alternative
interpretations of these two events as captures on $^{14}$N, with
binding energies consistent with the calculated value listed in
the table, for example \cite{ABC93},
$B_{\Xi^{-}}(^{14}{\rm N}) = 0.35 \pm 0.20~~{\rm MeV}$. Furthermore,
a likely interpretation of the third event \cite{NSY97} is due to
capture on $^{16}$O, with
$B_{\Xi^{-}}(^{16}{\rm O}) = 0.31 \pm 0.23~~{\rm MeV}$.
Clearly, whereas these emulsion events are consistent with capture from
$3D$ {\it atomic} states, they are useless as a source of
information regarding the $\Xi$-nucleus interaction.

One clearly needs an alternative source of information on the $\Xi$-nucleus
strong interaction. Such an alternative source is the measurement of
X-ray energies from transitions between low lying levels of $\Xi^-$
hadronic atoms. The experimental accuracies of the proposed measurements
are such that meaningful information on the $\Xi$-nucleus
interaction is likely to be obtained. In arguing the case for doing
experiments with $\Xi^-$ atoms, we follow the example of $\Sigma^-$
atoms which, as was recently shown \cite{BFG94},
gives rise to meaningful information
about the $\Sigma$ nucleus interaction, particularly in the absence of
any systematic evidence for bound $\Sigma$ hypernuclei.

Experiments with stopped $\Xi ^-$ hyperons had been proposed by
Zhu {\it et al.} \cite{ZDG91} and by Akaishi and collaborators \cite{KKA95}
in order to produce some of the lightest $\Lambda \Lambda$
hypernuclei, $_{\Lambda \Lambda}^{~~6}$He and
$_{\Lambda \Lambda}^{~~4}$H
(if the latter is particle stable) respectively, by looking for a peak
in the outgoing neutron spectrum in the two-body reaction

\begin{equation}
\label{eq:lamlam}
\Xi ^- \quad + \quad ^AZ \quad \longrightarrow
\quad _{\Lambda \Lambda}^A(Z-1) \quad + \quad n \quad .
\end{equation}
These proposals motivated the AGS experiment E885 \cite{May98} on $^{12}$C,
using a diamond target to stop the $\Xi ^-$ hyperons resulting from the
quasi-free peak of the $(K^-, K^+)$ initial reaction.
Finally, stopping $\Xi ^-$ hyperons in deuterium has been used in the AGS
experiment E813 to search for the doubly strange $H$ dibaryon through the
reaction $(\Xi ^- d)_{\it atom} \rightarrow Hn$, as reviewed recently by
Chrien \cite{Chr98}.

In Sec. \ref{sec:exp} we consider the experimental features of 
measuring X-ray transitions in $\Xi^-$ atoms,
 and in Sec. \ref{sec:tgts} we discuss possible target
nuclei, using a $t\rho$ optical potential. Alternative choices, using
density dependent optical potentials, are discussed in Sec. \ref{sec:DD}.
The special case of $\Xi^-$ atoms of deuterium, which is connected to recent
searches for the $H$ dibaryon, is discussed in Sec. \ref{sec:Xid}.  Section
\ref{sec:sum} summarizes the present work.

\section{Experimental considerations}
\label{sec:exp}

In considering the feasibility of measuring strong interaction
effects in $\Xi^-$ atoms, we are guided by the successful observation
and measurement of strong interaction
effects \cite{BBE75,BBB78,PEG93} in $\Sigma^-$ atoms.
The $\Xi^-$ and the $\Sigma^-$ hyperons have very similar masses and
lifetimes, namely 1321.32 and 1197.34 MeV for $\Xi^-$ and $\Sigma^-$
respectively, and 0.1642 and 0.1482 nsec for $\Xi^-$ and $\Sigma^-$
respectively. Consequently,
major differences between the experimental X-ray counting rates should result
mostly from differences between the
production mechanism of the two kinds of  hadronic atoms.
In the $\Sigma^-$ case the production mechanism is  the
$K^-p \longrightarrow \Sigma^- \pi^+$ reaction at rest. The outgoing
$\Sigma^-$ has a kinetic energy of 12.4 MeV which means that it stops
in a heavy target such as Pb or W in less than 10$^{-11}$ sec
and the outgoing pion has a kinetic energy of 83.2 MeV which makes it
quite easy to detect. For the $\Xi^-$ particle the situation is less
favourable. The production mechanism is the $K^-p \longrightarrow \Xi^-K^+$
reaction which has a total cross section larger than 100 $\mu$b only in the
kinetic energy range for $K^-$ of 1.0 to 1.8 GeV
(see results summarized in Ref. \cite{DGa83}). Consequently
the majority of the produced $\Xi^-$ will be in the energy range of
100 to 300 MeV, leading to stopping times in heavy targets of 0.01 to
0.3 nsec respectively. It is therefore essential that the $\Xi^-$
particles slow down in a heavy
degrader and not in the liquid hydrogen used for the production where the
stopping times are much longer. A laminar
target structure similar to that used in  $\Sigma^-$ experiments
\cite{PEG93,HEG88} seems the most suitable. Without detailed calculations
for specific target and degrader configurations
it is only possible to estimate that up to 50\% of the produced $\Xi^-$
may survive the slowing down process. Detection in coincidence with the
outgoing $K^+$ is a must in such an experiment. The energy of this
$K^+$ will be about 500 MeV.

When interested in the measurement of strong interaction effects
in hadronic atoms
it is also necessary to consider losses during the atomic cascade
process. Here, again, it is possible to make use of the fact that $\Sigma^-$
atoms have been successfully observed, and that the $\Xi^-$ and the
$\Sigma^-$ hyperons have very similar masses and lifetimes.
We have performed calculations of the atomic cascade for $\Sigma^-$ atoms
of Pb, in order to compare the
predicted absolute X-ray yields with experiment \cite{PEG93}.

The cascade program used here was based on that originally written by
H\"{u}fner \cite{Huf66}
for muonic atoms and later modified to include strong interaction
effects. The hadronic atom is assumed to be formed in a state of large radial
quantum number $n$.
To accommodate departures from a purely statistical distribution the initial
population used is of the form

\begin{equation}\label{eq:popdist}
p(l)=(2l+1)e^{\alpha l}
\end{equation}

\noindent
where for $\alpha=0$ the statistical distribution is obtained.
The atom initially de-excites by Auger transitions and later by radiative
transitions with the emission of X-rays. Finally in a state of low angular
momentum $l$ the hadron is absorbed by the nucleus and the X-ray cascade
terminates. The probabilities for Auger and radiative transitions are
calculated with the usual electromagnetic expressions \cite{EKe61}. For the
strong interaction the input to
the calculations are the measured or calculated absorption widths of low $n$
circular $(n, l=n-1)$ atomic states which are then scaled \cite{BWe95} to
obtain widths for non-circular states of a given $l$ and increasing $n$.
The probability that the hadron may decay during the  atomic
cascade is also included in the calculation.

\begin{figure}
\epsfig{file=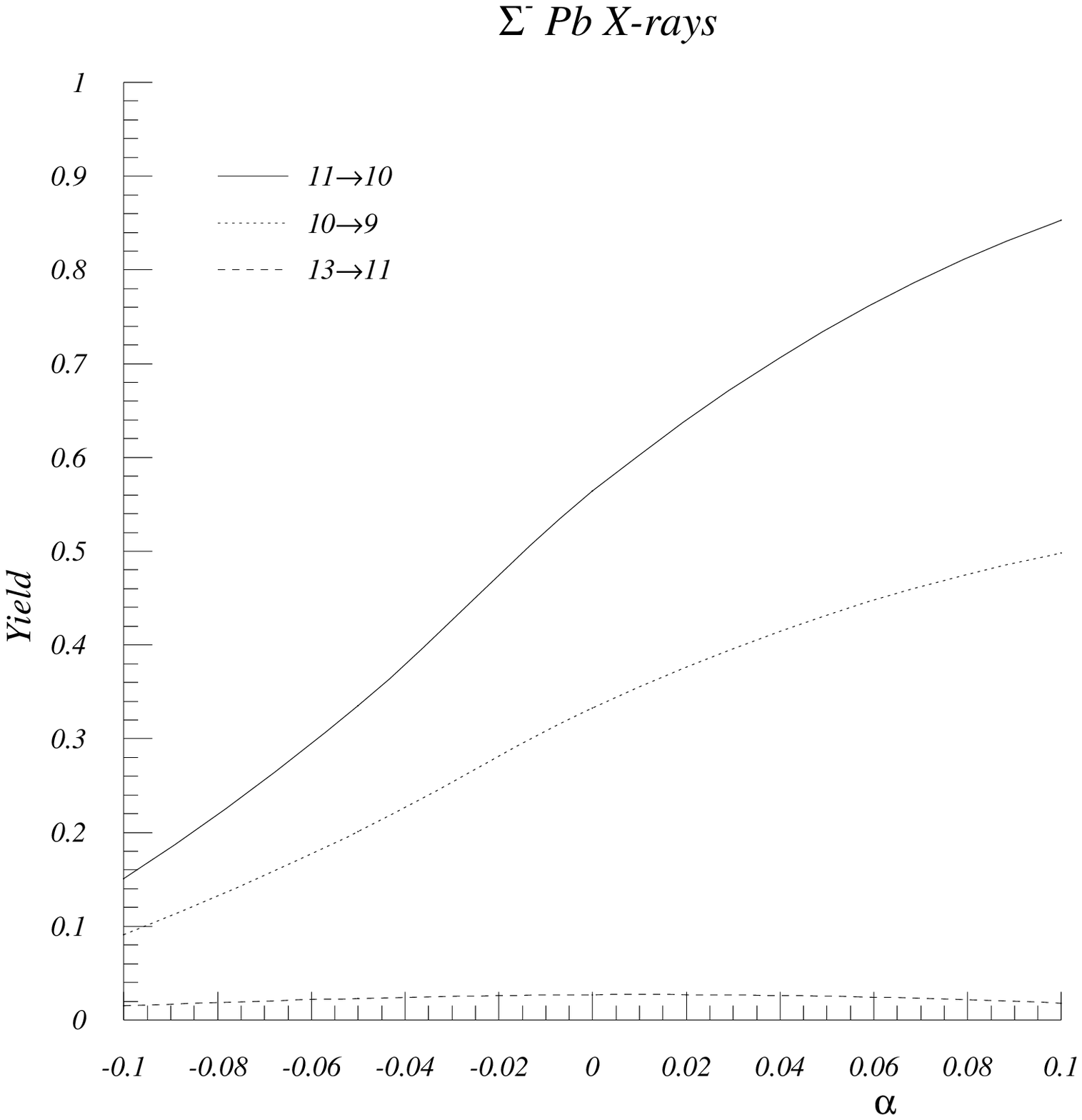,height=140mm,width=120mm,
bbllx=45,bblly=180,bburx=499,bbury=658}
\caption{Calculated absolute yields of X-ray transitions in $\Sigma^-$ Pb atoms
as a function of the initial population parameter $\alpha$. Results are
shown for $n_i \rightarrow n_f$ transitions with $l=n-1$. }
\label{fig:SigPbY}
\end{figure}

\begin{figure}
\epsfig{file=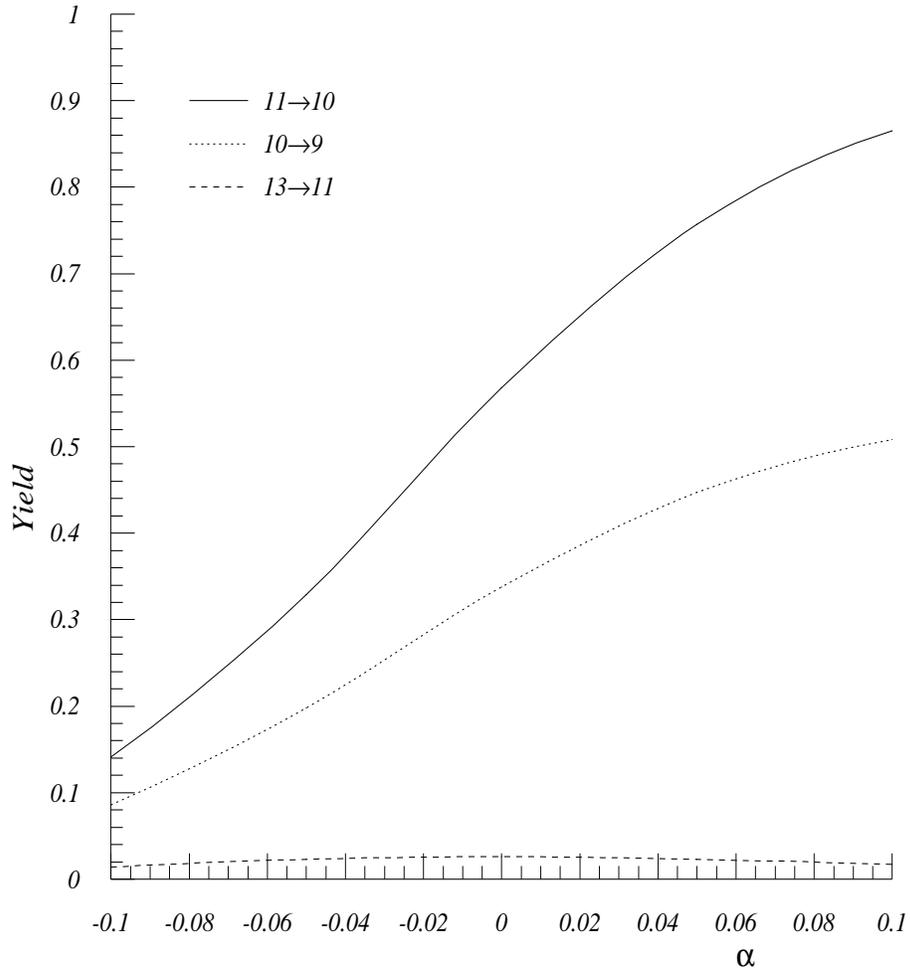,height=140mm,width=120mm,
bbllx=45,bblly=180,bburx=499,bbury=658}
\caption{Same as Fig. \protect{\ref {fig:SigPbY}} but for $\Xi^-$ Pb atoms.}
\label{fig:XiPbY}
\end{figure}

For the
strong interaction widths we used values predicted by the
density dependent $\Sigma$ nucleus potential
which reproduces all the   available data \cite{BFG94}. From a fit to the
measured relative X-ray yields \cite{PEG93}
we obtain the value $\alpha=-0.056\pm0.020$ which corresponds to a
small departure from a purely statistical distribution. The predicted
absolute yields for the three transitions of interest are shown in
Fig. \ref{fig:SigPbY} where it is seen that their
variation with $\alpha$ over its fitted range of uncertainty
is not  very large.
Note that the $11M \rightarrow 10L$ and
$10L \rightarrow 9K$ transitions in $\Sigma^-$ Pb have been observed
\cite{PEG93}.
Next we calculated the yields for $\Xi^-$ atoms of Pb for a range of
values of $\alpha$. Strong interaction
widths of the relevant levels were calculated with a $t\rho$ potential
with $b_0=0.25+i0.04$ fm as discussed below.
Figure \ref {fig:XiPbY} shows the calculated absolute
yields, which turn out to be very similar to the corresponding values for
the $\Sigma^-$ Pb atom. In both cases the $13 \rightarrow 11$
transition is also shown because its energy is very close to that of the
$10 \rightarrow 9$ transition and the two will have to be separated.
It is therefore concluded that $\Xi^-$ atoms could be observed
if the production yield and detection efficiency are not too small compared
to the corresponding values for $\Sigma^-$ atoms.

\section{Selection of targets}
\label{sec:tgts}
\subsection{Potentials and criteria}
\label{subsec:potls}

When selecting targets for  possible experiments on $\Xi^-$ atoms,
it must be assumed that such experiments will probably
not be feasible
on more than very few targets, and one must therefore ask
whether it is at all likely that  useful information on the
interaction of $\Xi^-$ with nuclei
will be obtained from the resulting rather limited range of data.
Useful hadronic-atom data normally consist of the strong-interaction
shift and width for the `last' level observed plus, occasionally, the
relative yield for the `upper' level, as discussed below. It has been
shown very recently \cite{Fri98} that the main features of the interaction
of $K^-$ and $\Sigma^-$ with nuclei, as found from analyses of all the
available data, may in fact, be obtained by analyzing
a small fraction of the available hadronic atom data, if the target
nuclei are carefully selected. A key point here  is to have target nuclei over
as wide a range of the periodic table as possible. This observation suggests
that  experiments on $\Xi^-$ atoms may provide useful information.

In order to have some idea of the expected strong interaction effects
in $\Xi^-$ atoms and on the expected yields of X-ray transitions,
it is necessary to adopt
some form of an optical potential which will describe, at least
approximately, the interaction of the $\Xi^-$ hyperon with nuclei and
its dependence on the nuclear mass. The so-called $t\rho$ potential
\cite{BFG97} in its simplest form is given by

\begin{equation}
\label{eq:trho}
2\mu V_{opt}(r) =
 -4\pi(1+\frac{\mu}{M}) b_0 \rho(r)
\end{equation}

\noindent
where $\mu$ is the $\Xi$-nucleus reduced mass, $M$ is the mass
of the nucleon and $\rho(r)$ is the nuclear density, normalized to
the mass number $A$.
The parameter $b_0$ is a complex parameter, and is usually related
to the hadron-nucleon scattering length \cite{BFG97}.
More refined  potentials have been used with other
hadronic atoms and will be discussed in the following section.
 For the present purpose we select the value
of Re$b_0=0.25$ fm which yields a potential depth of about 20 MeV inside
nuclei, and Im$b_0=0.04$ fm, yielding for the imaginary potential a depth
of about 3 MeV. Whereas the real potential may be regarded as `typical',
according to the discussion in Sec. \ref{sec:int},
 the imaginary potential is
about twice as large as estimated \cite{YMH94} in model D. Reducing the
imaginary potential  will only cause
the calculated widths of the states to decrease by roughly the same proportion,
 and the  relative yields (see below) of transitions to become larger.
This will not, however, change the last observed atomic level. A further
comment on the choice of the imaginary potential is made in
Sec. \ref{sec:DD}.

In choosing criteria for the suitability of a transition as a source of
information on the $\Xi$ nucleus interaction, we are guided by  experience
with other hadronic atoms \cite{BFG97,Bat82} and select X-ray transitions
$(n+1,l+1) \rightarrow (n,l)$ between circular atomic states $(n=l+1)$
with energies greater than 100 keV, where the strong interaction
shift for the `last' $(n,l)$ level
is at least 0.5 keV and the width less than about 10 keV.
The `upper' level relative yield, defined as the ratio of the intensity of the
$(n+1,l+1) \rightarrow (n,l)$ X-ray transition to the summed intensity of
all X-ray transitions feeding the $(n+1,l+1)$ state, is also required
to be at least 10\%.
The `upper' level relative yield \cite{Bat82} is given by the
ratio of the
radiative width  for the $(n+1,l+1) \rightarrow (n,l)$ transition
to the total width of the $(n+1,l+1)$ level feeding the $(n,l)$ level.
A measurement of the relative yield enables the strong interaction width of
the `upper' $(n+1,l+1)$ level to be determined, in addition to
deducing the strong-interaction level shift and width for the `last'
$(n,l)$ level from the X-ray transition energy discussed above.
Altogether, these are the strong-interaction data provided by
measuring atomic X-ray spectra.

\subsection{Results}
\label{subsec:results}
\begin{figure}
\epsfig{file=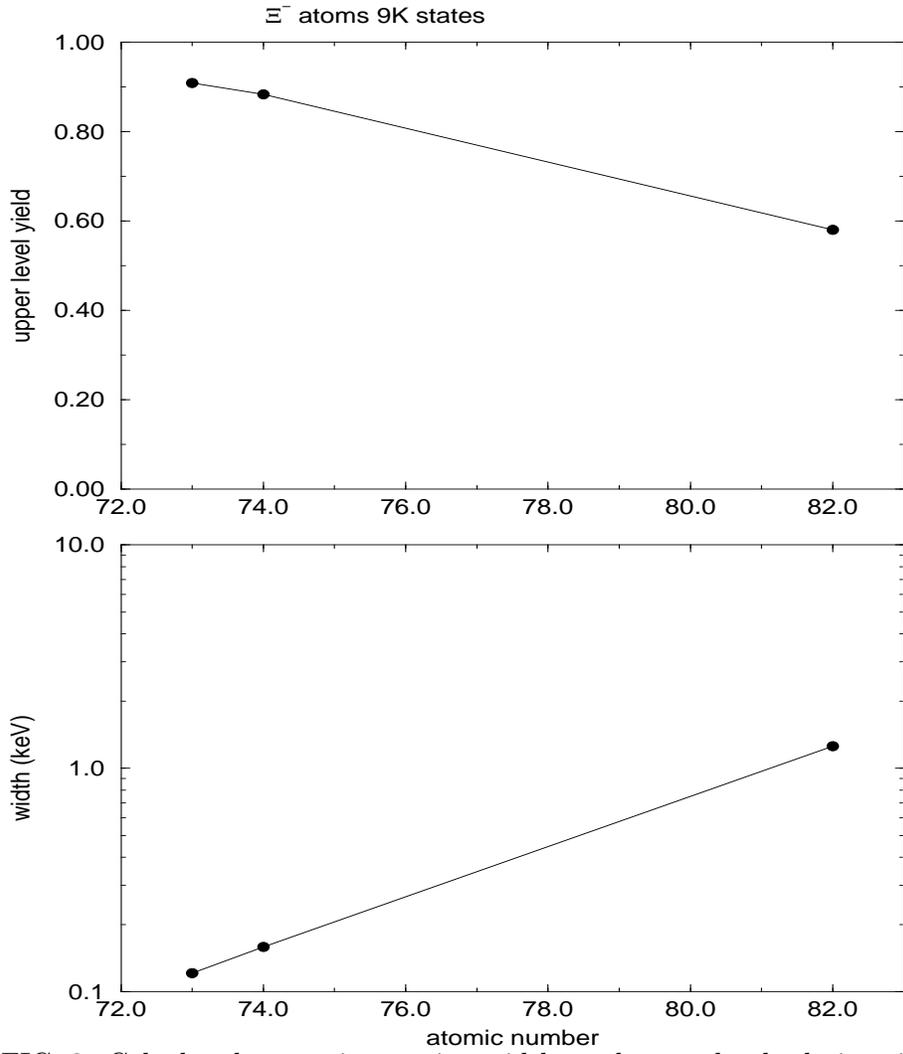,height=140mm,width=120mm,
bbllx=40,bblly=54,bburx=538,bbury=766}
\caption{Calculated strong interaction widths and upper level relative
yields for the $9K$ level in heavy $\Xi^-$ atoms.}
\label{fig:XiPb}
\end{figure}

\begin{figure}
\epsfig{file=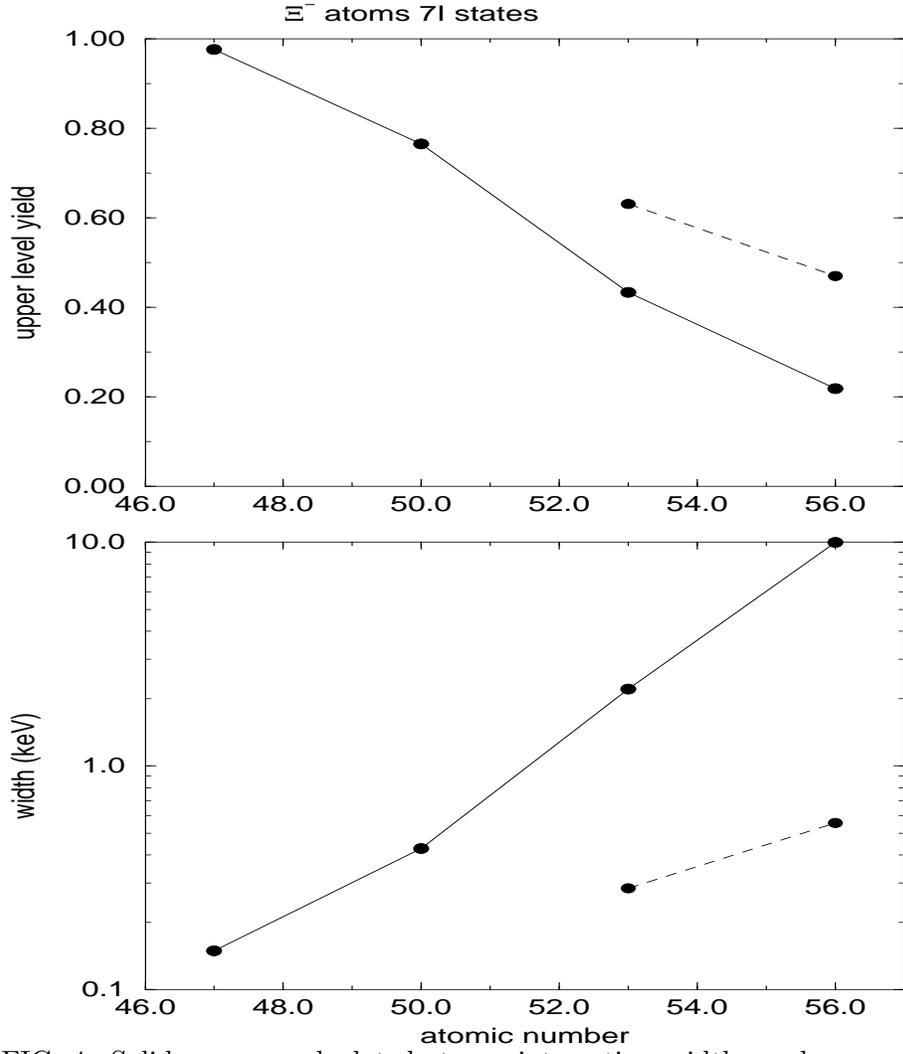,height=140mm,width=120mm,
bbllx=69,bblly=54,bburx=480,bbury=765}
\caption{Solid curves: calculated strong interaction widths and upper level relative
yields for the $7I$ level in medium-heavy $\Xi^-$ atoms.
The dashed curves are for $b_0=-0.25+i0.04 $ fm, i.e. a repulsive real
potential.}
\label{fig:XiSn}
\end{figure}

Strong interaction shifts and widths of $\Xi^-$ atomic levels have been
calculated using the above optical potential for a large number of nuclei.
As the overlap
of atomic wavefunctions with nuclei vary smoothly with charge number, it
is to be expected that generally shifts, widths and yields will vary smoothly
along the periodic table.
Figure \ref{fig:XiPb} shows calculated widths and `upper'
level relative yields for the
$9K$ level in $\Xi^-$ atoms for several heavy targets. Calculated
shifts (not shown) for the chosen potential are generally equal to or up to
30\% larger than the widths, as long as both are smaller than 10 keV. It
is seen from the figure that  Pb  is a suitable target for the
experiment.
Figure \ref{fig:XiSn} shows similar results for the $7I$ state in medium-heavy
$\Xi^-$ atoms and it is seen that  a suitable target may be found
near Sn or I.
 The dashed lines in this figure
are obtained by reversing the sign of the real potential used for calculating
the solid curves. It is seen that in such a case the range of
suitable targets will move to between I and  Ba,
where the strong interaction
width and relative yield are more acceptable.
The sign of the strong interaction shift will be reversed in this case, but
it has no experimental consequences.
This exemplifies a general property of hadronic atoms, which are dominated
by the Coulomb interaction, namely, that large variations in the strong
interaction potential will move the proposed targets only a few units of
charge along the periodic table.

\begin{figure}
\epsfig{file=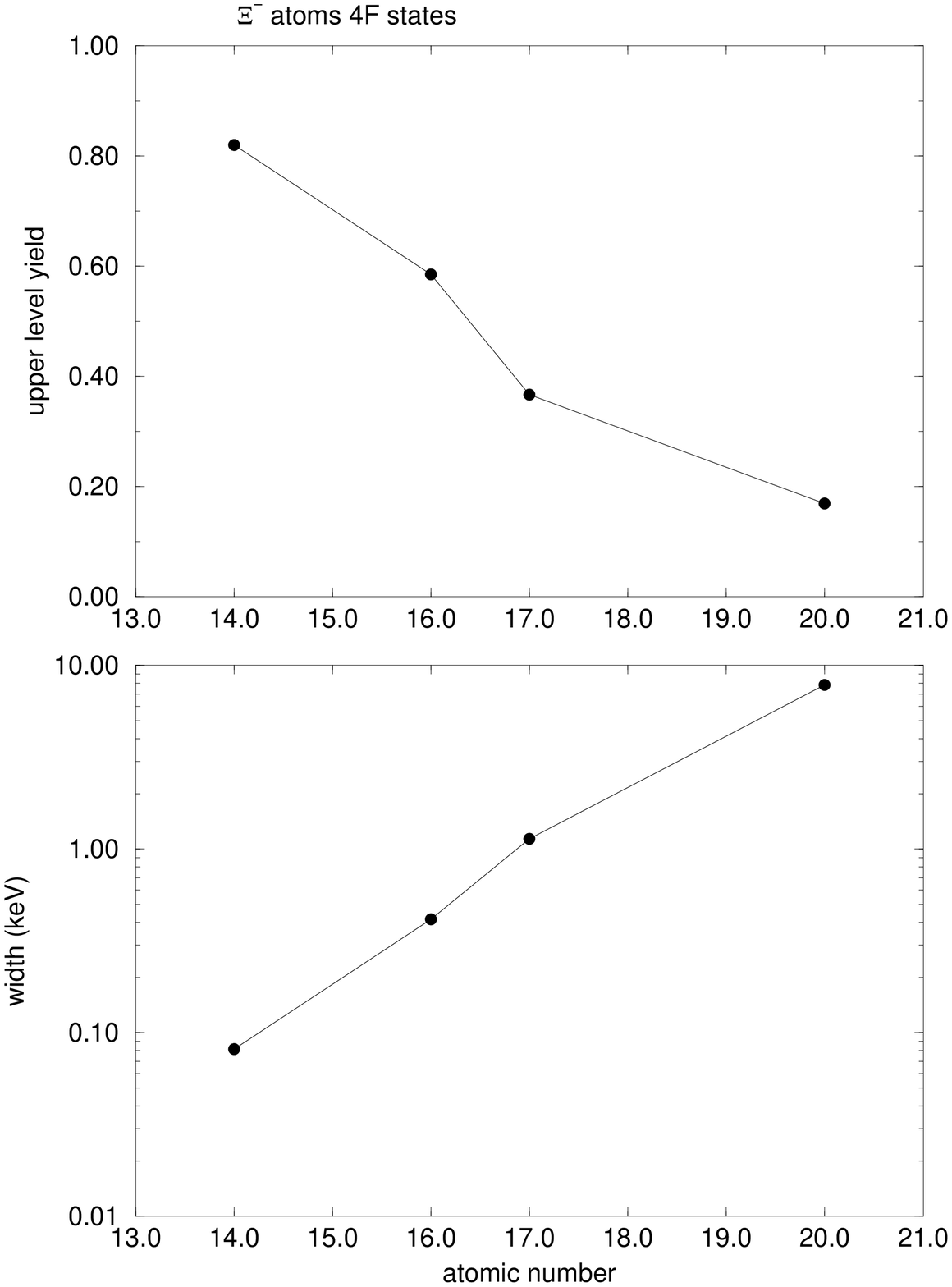,height=140mm,width=120mm,
bbllx=29,bblly=54,bburx=551,bbury=765}
\caption{Calculated strong interaction widths and upper level relative
yields for the $4F$ level in $\Xi^-$ atoms.}
\label{fig:XiSi}
\end{figure}

\begin{figure}
\epsfig{file=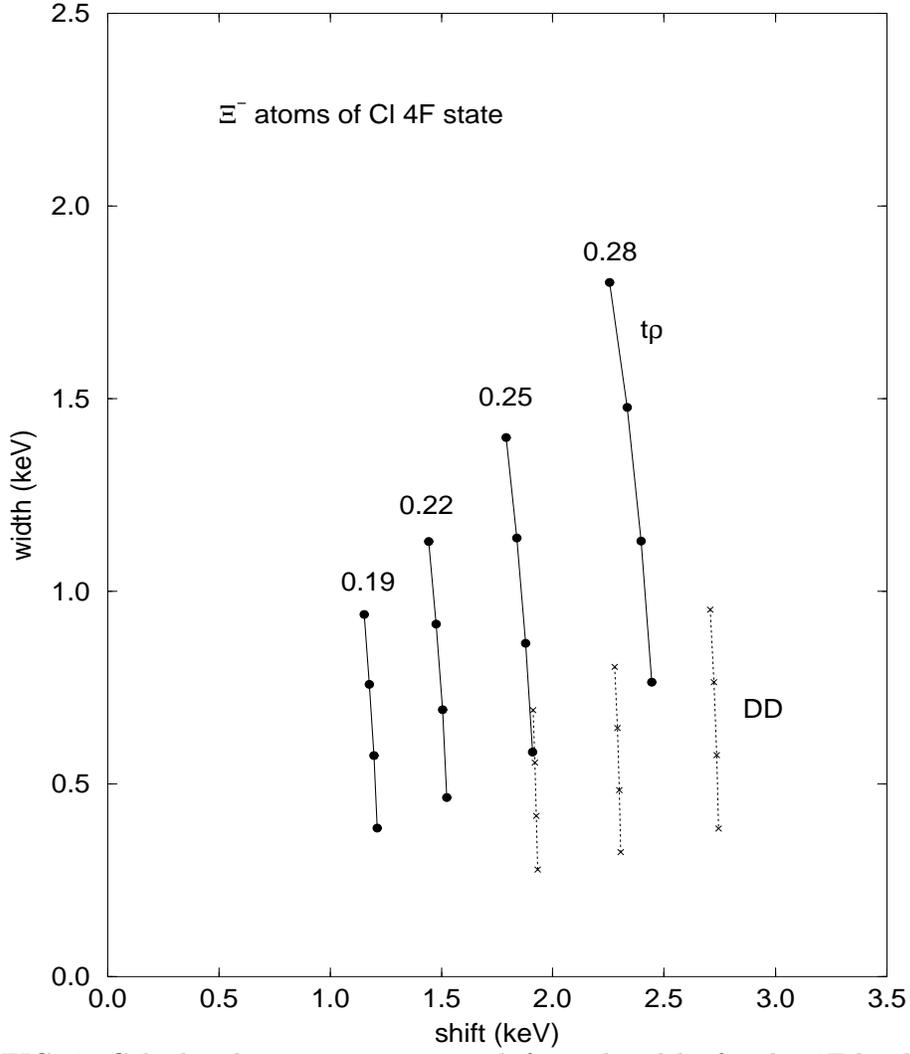,height=140mm,width=120mm,
bbllx=51,bblly=92,bburx=518,bbury=670}
\caption{Calculated strong interaction shifts and widths for the $4F$
level in $\Xi^-$ atoms of Cl for different optical potentials, see text
for detail.}
\label{fig:XiClall}
\end{figure}

Figure \ref{fig:XiSi} shows results for the $4F$ state of $\Xi^-$ atoms,
where it is seen that for a Si target the effects could be too small to
measure whereas for Ca the width could be too large and the relative yield too
small. In this region a Cl target may be appropriate, perhaps in the
form of the liquid CCl$_4$.
More detailed results for Cl are shown in Fig. \ref{fig:XiClall} where the
sensitivities to assumptions regarding the optical potential are also
typical of results for other targets.
The solid curves connect points obtained
within the $t\rho$ potential Eq. (\ref{eq:trho})
for  fixed values of Re$b_0$, listed above the lines. The four points
along each line correspond to values of Im$b_0$ from 0.05 fm
down to 0.02 fm in steps of 0.01 fm.
Departures from this $t\rho$ potential are represented by the dotted lines,
calculated from phenomenological density dependent (DD)
real potentials similar to those
found from analyses of experimental results for $\Sigma^-$ and $K^-$
atoms \cite{BFG97}. The imaginary part of the potential is of the $t\rho$
type and the points along the dotted lines correspond to the same
values of Im$b_0$ as above. The real potentials in these calculations
are similar to the real potential for $\Sigma^-$ atoms \cite{BFG97},
having an attractive pocket about 5-10 MeV deep outside the nuclear surface,
with a repulsive potential of about 20-30 MeV in the nuclear interior.
The results in the figure serve only to
illustrate the expected range
of strong interaction effects.
If the actual values of shift and width turn out to be within the area covered
by the lines, these effects will most likely be measurable.

\begin{figure}
\epsfig{file=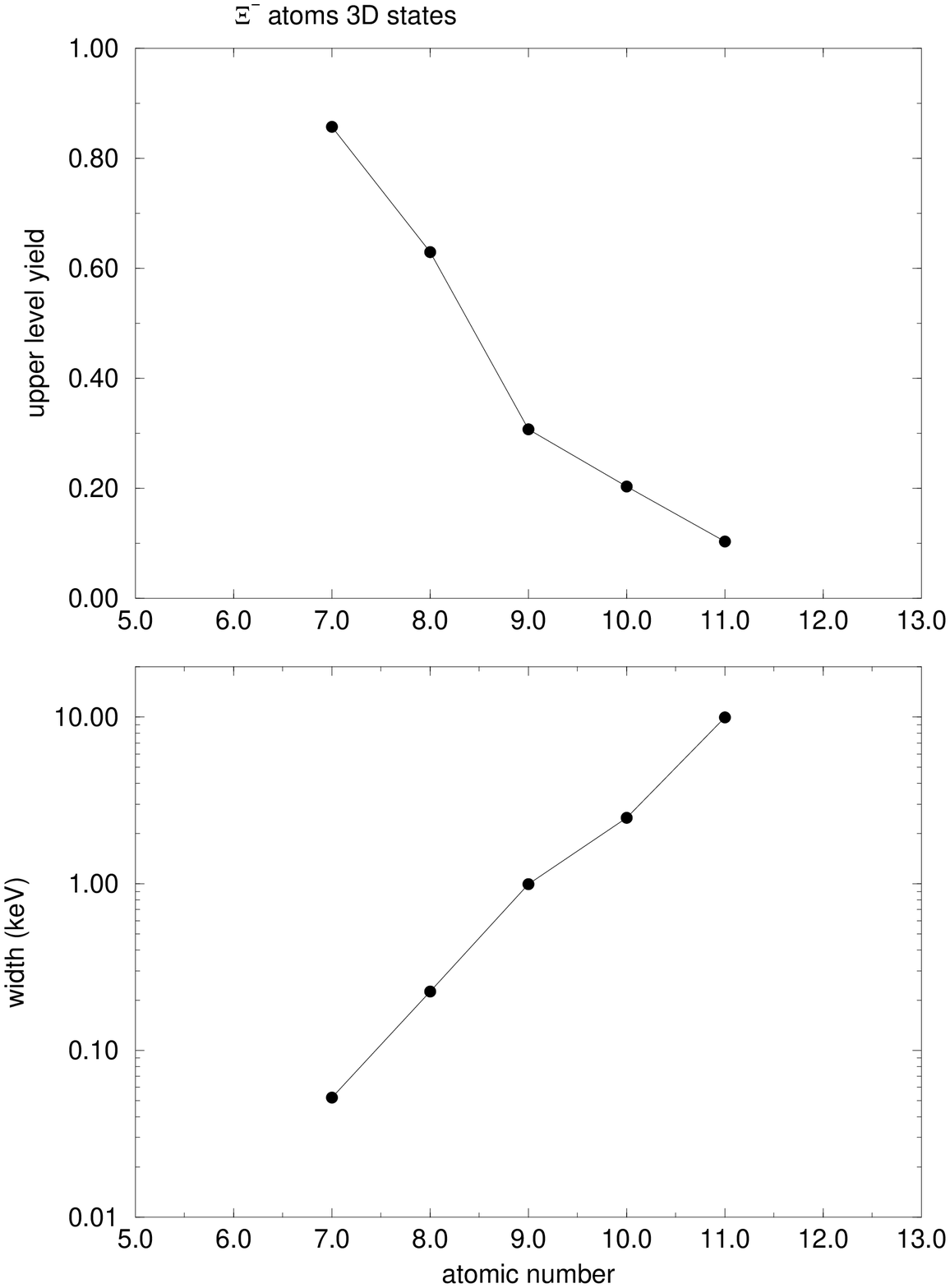,height=140mm,width=120mm,
bbllx=29,bblly=54,bburx=551,bbury=766}
\caption{Calculated strong interaction widths and upper level relative
yields for the $3D$ level in $\Xi^-$ atoms.}
\label{fig:XiF}
\end{figure}

\begin{figure}
\epsfig{file=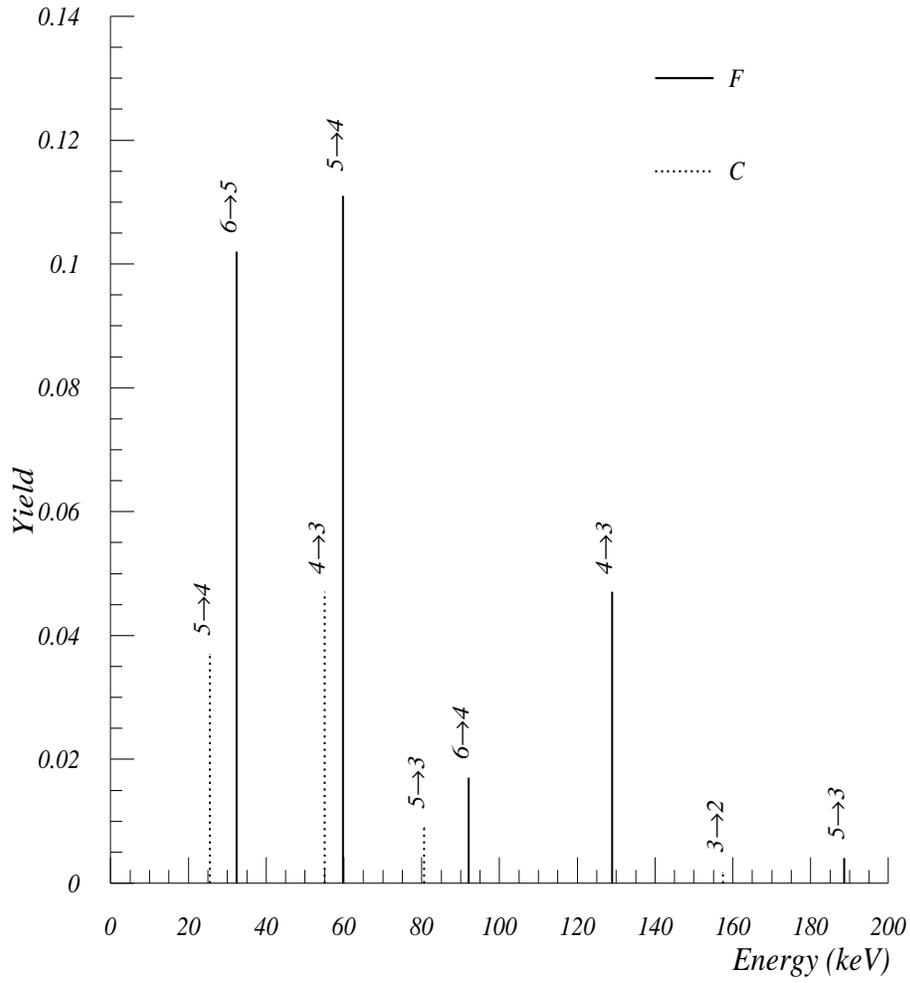,height=140mm,width=120mm,
bbllx=45,bblly=181,bburx=501,bbury=658}
\caption{Calculated X-ray spectrum for a teflon target.}
\label{fig:XiCF2}
\end{figure}

Figure \ref{fig:XiF} shows calculated results for $3D$ states in several
$\Xi^-$ atoms, and it is seen that F may be a suitable target, possibly
in the form of teflon (CF$_2$). To see what could be the effects of the carbon
in the target we have also performed
full atomic cascade calculations for C and F,
and the expected X-ray spectrum for teflon
is shown in Fig. \ref{fig:XiCF2}. It seems
that the presence of carbon in the target should not affect the possibility
of observing transitions in $\Xi^-$ F atoms.

Results for the $t\rho$ potential with $b_0=0.25+i0.04$ fm are summarized
in Table \ref{tab:preds}, which includes several
suitable targets across the periodic
table where strong interaction effects are likely to be measurable.
A range of targets is necessary if unusual features of the interaction, such
as density dependence beyond the $t\rho$ prescription,
are to be unraveled by the experiment.

\section{$\Xi$ NUCLEUS DENSITY DEPENDENT POTENTIALS}
\label{sec:DD}

In the previous section we have discussed $\Xi^{-}$ X-ray spectra
across the periodic table, assuming a $\Xi$-nucleus strong
interaction potential of the form $V^{(\Xi)}=t\rho$, Eq.
(\ref{eq:trho}). Specific predictions were given, using the value
$b_{0} = 0.25 + i 0.04$ fm, and the sensitivity to departures from
this value was displayed in Fig. \ref{fig:XiSn} and \ref{fig:XiClall}.
Conversely, one
may explore the extent to which measuring, say, four X-ray spectra
(e.g. Table \ref{tab:preds}) will determine the $\Xi$-nucleus potential.
In the present section we address this question  by
choosing a one-boson-exchange (OBE) motivated  $\Xi$-nucleus
potential, use it to
evaluate the strong-interaction shifts and widths
for several  targets across the periodic table, and then
solve the inverse problem, namely determine the best-fit
parameters for  $\Xi$ nucleus  potentials of different forms.

\begin{figure}
\epsfig{file=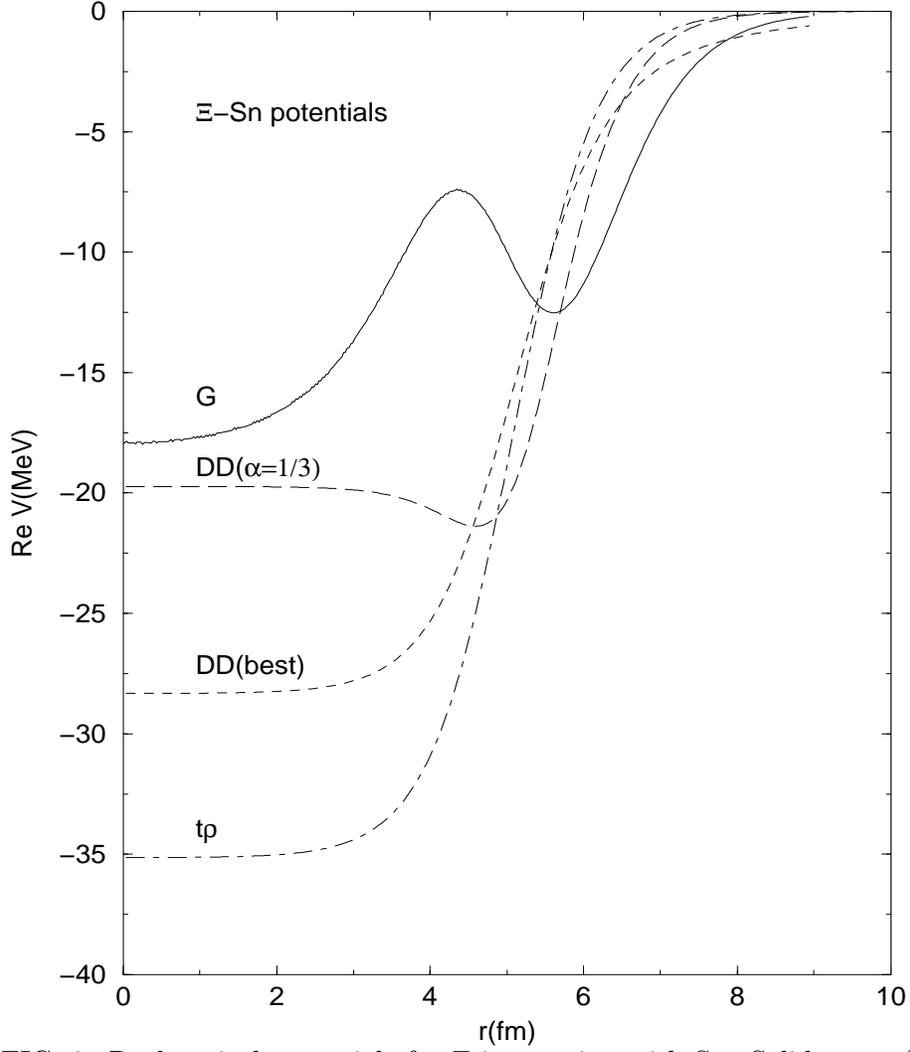,height=140mm,width=120mm,
bbllx=42,bblly=92,bburx=516,bbury=670}
\caption{Real optical potentials for $\Xi$ interaction with Sn.
Solid curve (G) is for Eq. (\protect{\ref{eq:fold}}) used to generate the
`pseudo data'. The other curves are the fitted potentials
obtained under various constraints. See text.}
\label{fig:VXiNuc}
\end{figure}

We start with the $\Xi N$ DD $G$-matrix interaction YNG derived by
Yamamoto {\it{et al.}} \cite{YMH94} from the Nijmegen OBE potential
model D \cite{NRD77}:

\begin{equation}\label{eq:gxin}
v_{\Xi N}\left( {r;k_F} \right)=\sum\limits_{i=1}^3
{\left( {a_i+b_ik_F+c_ik_F^2} \right)\exp \left( {{{-r^2}
\mathord{\left/ {\vphantom {{-r^2} {\beta _i^2}}} \right.
\kern-\nulldelimiterspace} {\beta _i^2}}} \right)}\   ,
\end{equation}
where the nuclear Fermi momentum $k_{F}$ is expressed in terms of the
local nuclear density $\rho$, $k_F=\left( {{{3\pi ^2\rho } \mathord{\left/
{\vphantom {{3\pi ^2\rho } 2}} \right. \kern-\nulldelimiterspace} 2}}
\right)^{{1 \mathord{\left/ {\vphantom {1 3}} \right. \kern-\nulldelimiterspace}
 3}}$. To simplify matters, while retaining a substantial part of the
 density dependence of $v_{\Xi N}$, we kept only the $a$ and $b$ terms.
 These $G$-matrix parameters are listed in Table \ref{tab:gxin} for the
 choice $r_{c}$ = 0.46 fm, which belongs to the range of values
 considered in Ref. \cite{YMH94}. We note that the $G$ matrix depends
 strongly on the hard-core radius $r_{c}$ chosen in the $\Xi N$ sector
 of model D. The resulting $v_{\Xi N}$ is repulsive at short $\Xi N $
 distances and attractive at distances $r \geq 0.8$ fm. The $\Xi$
 nucleus DD real potential, obtained by folding $v_{\Xi N}$ with the
 nuclear density $\rho$:

 \begin{equation}
 \label{eq:fold}
V^{\left( \Xi  \right)}\left( {r;\rho } \right)=\int {v_{\Xi N}\left( {\left|
{{\bf r}-{\bf r'}} \right|;k_F\left( \rho  \right)} \right)\rho \left( {r'}
\right)d^3r'}\   ,
 \end{equation}
is shown in Fig. \ref{fig:VXiNuc} (G, solid line) for the nucleus Sn.
Its depth is about 18 MeV (attractive) in medium-weight and heavy
nuclei  and about half  of that in light nuclei,
so it falls conveniently
within the range of values considered as `realistic' in Sec.
\ref{sec:int}. However, as is clear by comparing the shape of this
$V^{(\Xi)}$ (potential G) with the (fitted, see below)
$t \rho$ potential, plotted as a dot-dash line in
the figure,   potential G extends considerably
further out than does the nuclear density distribution $\rho$ . The
difference in the r.m.s. radius values amounts to about 1.3 fm. The
main contribution to this difference does not arise just from folding
the finite-size $v_{\Xi N}$ into $\rho$, but is due to the summed
effect of the DD terms `$b$' which produce repulsion on top of
the attractive contribution due to the sum of the `$a$' terms.
Retaining the `$c$' terms in Eq. (\ref{eq:gxin}) would have reduced
the r.m.s. radius of potential G, but in addition it would have
increased the nuclear-matter depth to about 30 MeV, a value which
currently \cite{Fuk98} is considered to be a gross overestimate.

For the imaginary part of $V^{(\Xi)}$ we retained, for simplicitly,
the $t \rho$ form with Im$b_{0} = 0.04$ fm as before.
We checked that replacing it by the corresponding
 DD form  \cite{YMH94}, as per Eqs. (\ref{eq:gxin},\ref{eq:fold}),
has a marginal effect on our results.
The reason is that since the real part of $V^{(\Xi)}$
is substantially stronger than the imaginary part, the
main effect of the density dependence is due to the real part of
$V^{(\Xi)}$.

The calculated strong-interaction shifts and widths, using potential G,
 are shown in Table
\ref{tab:tgts} for several target nuclei, including the five targets
listed in Table \ref{tab:preds} which were motivated by the $t \rho$
potential with a similar depth. Where comparison can be made, these
shifts and widths are larger by an order of magnitude than those in
Table \ref{tab:preds}, due to the greater extension of potential G
which pulls in the atomic wavefunctions by up to 2 fm (judging by the
inward shift of the position of the
maximum of the atomic wavefunctions at about 20 fm).
Consequently, with most of the calculated shifts and widths
exceeding 10 keV, the targets (F, Cl, I, Pb) will no longer be the
optimal ones for experiments with $\Xi^{-}$ atoms. As discussed
in Sec. \ref{subsec:results}, one can identify neighbouring targets by
going several charge units down in each sequence to reestablish
optimum conditions. Indeed, according to Table \ref{tab:tgts}, an
optimal set of targets could consist of (O, S, Sn, W).

For testing the inversion procedure under the extreme conditions
set up by using potential G,
the shifts and widths calculated for these four $\Xi^{-}$ atoms were
considered as pseudo data. Standard measurement errors of $20\%$
were ascribed, and randomized values of these pseudo data quantities
were then derived and used for fitting a given form of the real potential
$V^{(\Xi)}$ (the imaginary part was always taken as $t \rho$). Two forms
were used:

(i) The  $t \rho$ form, but with $b_0$ determined from the fit.
 The best fit parameter $b_0$ was found to be

\begin{eqnarray}
{\rm Re} b_{0}= 0.402 \pm 0.022~~{\rm fm},\quad\quad
 {\rm Im} b_{0}= 0.053 \pm 0.021~~{\rm fm}, \nonumber
\end{eqnarray}
with $\chi ^{2}/F$, the $\chi^2$  of the fit per degree of freedom,
assuming a relatively high value of $\chi ^{2}/F = 8.8$. The real part
of this potential is shown in Fig. \ref{fig:VXiNuc}. Note that it is
twice as deep in the nuclear interior as
 potential G, owing to the analytic continuation from the surface
region where the {\it fitted} potential must assume larger values
than before.
However, with such a distinct difference between the geometries
of the two potentials, the quality of fit produced by the $t \rho$
best-fit potential to the  pseudo data due to
 potential G is rather poor.

(ii) A phenomenological
DD potential of the form Eq. (\ref{eq:trho}), where the parameter
$b_0$ now depends on density:

\begin{equation}\label{eq:dd}
 b_{0}(\rho) = b_{0}+ B_{0}[\rho (r)/\rho (0)]^\alpha.
\end{equation}
For the imaginary part, the $t \rho$ form Eq. (\ref{eq:trho}) was
maintained, i.e. Im$B_{0} = 0$. If the exponent $\alpha$ is taken as
$\alpha = 1/3$, in order to simulate the density dependence of the
underlying $G$-matrix $v_{\Xi N}$ (Eq. (\ref{eq:gxin})), then the
best-fit parameters are

\begin{eqnarray}
{\rm Re} b_{0}&=&1.27 \pm 0.21 ~~{\rm fm},\quad\quad
   {\rm Im} b_{0}=0.019 \pm 0.009 ~~{\rm fm}, \nonumber \\
{\rm Re} B_{0}&=&-1.05 \pm 0.26 ~~{\rm fm}, \quad\quad \chi^{2}/F = 4.65. \nonumber
\end{eqnarray}
This potential is similar in depth to  potential G,
its r.m.s. radius is substantially larger than that of the $t \rho$ potential,
and as
$\rho \rightarrow 0$ it is about 3 times stronger than the best-fit
$t \rho$ potential discussed above. The quality of this fit is
quite reasonable.

If $\alpha$ is allowed to vary, it is found that the value
$\alpha = -0.7$ provides a very good fit to the  pseudo data:

\begin{eqnarray}
{\rm Re}b_{0}&=&0.268 \pm 0.033 ~~{\rm fm},  \quad\quad
  {\rm Im}b_{0}=0.040 \pm 0.009 ~~{\rm fm}, \nonumber \\
{\rm Re}B_{0}&=&0.056 \pm 0.011 ~~{\rm fm}, \quad\quad \chi^{2}/F = 1.77. \nonumber
\end{eqnarray}
This best-fit phenomenological DD potential is also plotted in
Fig. \ref{fig:VXiNuc}.
It is deeper within the nucleus than  potential G, becoming
less attractive than the latter in the immediate surface region outside
the nucleus, and then (since $\alpha$ is negative) becoming more
attractive in the more extreme surface region (8 fm and beyond, for Sn).

It is clear that neither the oversimplified $t \rho$ potential, nor the
phenomenological DD potentials defined by Eq. (\ref{eq:dd}), are capable of
reproducing the oscillatory behaviour of  potential G
in the nuclear surface region. The best-fit phenomenological
DD potential probably gives a
 reliable indication of the overall attractive nature of the
underlying potential $V^{(\Xi)}$, both yielding similar strengths of
about 10 MeV in the nuclear surface region.
It is noteworthy that the fitted imaginary part is practically the same for
both potentials, probably due to the same $t \rho$ form assumed
in the two cases.

\section{$\Xi^-$ deuterium atoms}
\label{sec:Xid}

In the previous sections we have discussed the possible observation of
X-rays from $\Xi^-$ atoms formed in nuclear targets and the possibility of
obtaining information about strong interaction effects in $\Xi^-$ atoms.
In the present section we discuss the properties of $\Xi^-d$ atoms and in
particular the formation of the doubly strange $H$ dibaryon through the reaction
$(\Xi^-d)_{atom} \rightarrow Hn$. An experiment to search for
this reaction is
in progress \cite{E813}. The formation of the $H$ particle is detected by
observation of the monoenergetic neutron, whilst the $\Xi^-$ is tagged by
the observation of the $K^+$ from the production reaction
$K^-p \longrightarrow \Xi^-K^+$ as discussed earlier in Sec.
\ref{sec:exp}.
The probability for $H$ dibaryon formation from $\Xi^-d$ atoms has been
calculated by Aerts and Dover \cite{ADo84} for reactions involving
$S$ and $P$ states
of the $\Xi^-d$ atom as a function of the mass of the $H$ particle.
(See Table II and Fig. 4 of Ref.\cite{ADo84}). Here we are specifically
concerned with calculating the probability for the $\Xi^-$ to interact with
the deuteron from an atomic $S$ or $P$ state in the $\Xi^-d$ atom.

For the $\Xi^-d$ atom, the atomic cascade proceeds generally as described in
Sec. \ref{sec:exp} but with an additional complication due to the presence
of Stark mixing \cite{DSS59}. This gives rise to transitions of the type
$(n,l\pm 1) \leftrightarrow (n,l)$ which increases the probability of
$\Xi^-d$ interactions from atomic $S$ states at large values of $n$. Since the
Stark mixing is proportional to the collision rate, and hence target density,
the effects are largest in liquid hydrogen. As the interactions take place at
high $n$, the lifetime of the $\Xi^-d$ atom is reduced
by the Stark mixing and the probability
for the $\Xi^-$ to decay during the atomic cascade decreases.

Some of the early qualitative discussions of Stark mixing \cite{DSS59}
for $K^-p$ atoms were
placed on a more quantitative basis by Leon and Bethe \cite{LBe62}. Their
calculations were later modified and extended by Borie and Leon
\cite{BLe80} to a range of exotic-hydrogen atoms such as
$\mu^-p,\ \pi^-p,\ K^-p$ and $\bar{p}p$.
Of interest for the present work, calculations
using the Borie and Leon model have been used to fit measured X-ray yields for
K- and L-series X-rays from antiprotonic hydrogen ($\bar{p}p$)
atoms \cite{Bat89}.
In using this model it has generally been found that, because of
approximations in the calculation of the absolute value of the Stark mixing
rate, an overall normalisation factor $k_{STK}$ is required to get a good
fit to the experimental X-ray data. Typical values of $k_{STK}$ required to fit
data for $\bar{p}p$ atoms are in the range 1.0 to 2.0.
In the present work
calculations are made for $k_{STK}$ over the range 0.5 to 5.0.

The present calculations were carried out using the Borie and Leon model
\cite{BLe80} with the program used in \cite{Bat89} for calculations on
antiprotonic hydrogen atoms, modified for $\Xi^-d$ atoms and including decays
of the $\Xi^-$. The principal input values are the Stark mixing normalization
factor $k_{STK}$ and the strong interaction widths $\Gamma_{1S}$ and
$\Gamma_{2P}$ for the $1S$ and $2P$ levels respectively. The strong interaction
width of the $3D$ and higher $(n,l = n-1)$-states were all set to zero. The
kinetic energy
of the $\Xi^-d$ atom was taken to be 1~eV \cite{BLe80}.

\begin{figure}
\epsfig{file=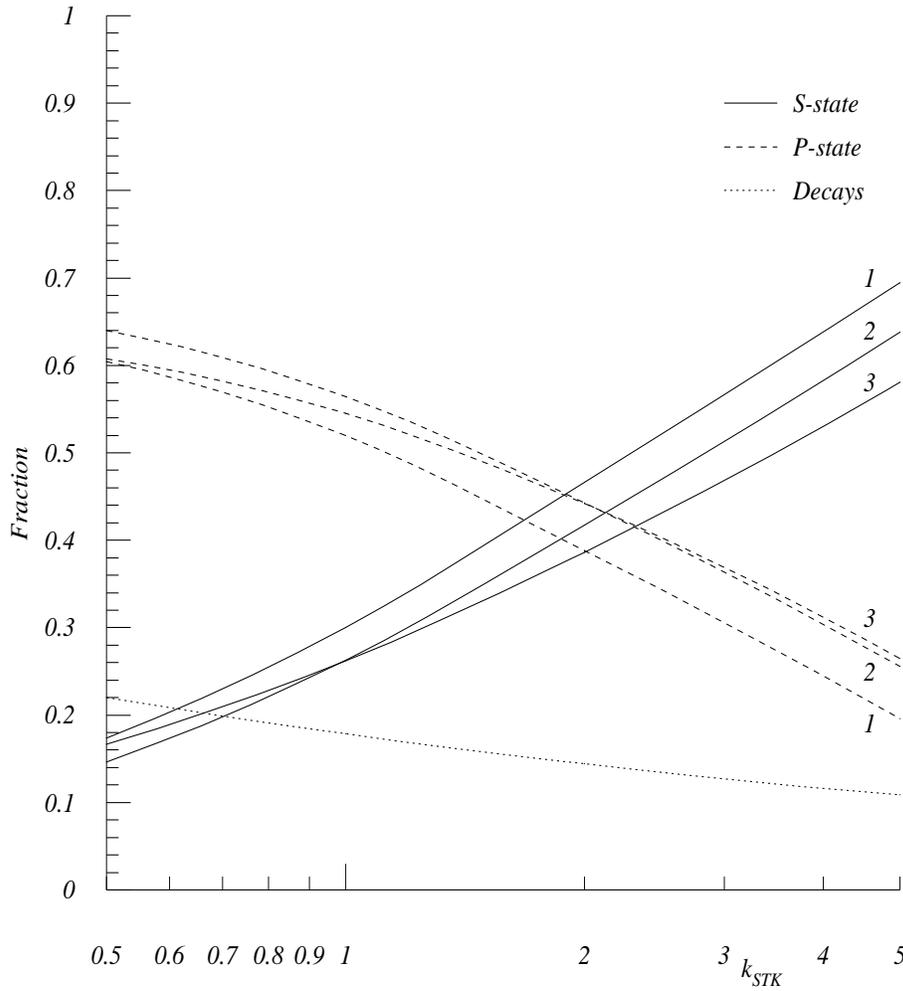,height=140mm,width=120mm,
bbllx=48,bblly=179,bburx=495,bbury=652}
\caption{Calculated fractions of $S$ and $P$ state capture, and of $\Xi^-$
decays, for three choices of the strong interaction parameters
$\Gamma_{1S}$ and $\Gamma_{2P}$ (see text) as a function of the Stark mixing
parameter $k_{STK}$. }
\label{fig:Xid10}
\end{figure}

The total strong interaction widths calculated by Aerts and Dover \cite{ADo84}
lie in the range $\Gamma_{1S}$ = 1 to 50~keV for $S$ states and $\Gamma_{2P}$ =
0.2 to 0.8~eV for $P$ states. In Fig. \ref{fig:Xid10} the fraction of $S$ and
$P$ state reactions as a function of the Stark mixing parameter $k_{STK}$
is shown for three sets of strong interaction parameters,
$\Gamma_{1S}$ = 10~keV, $\Gamma_{2P}$ = 0.2~eV (labelled 1),
$\Gamma_{1S}$ = 10~keV, $\Gamma_{2P}$ = 0.8~eV (labelled 2) and
$\Gamma_{1S}$ = 1~keV, $\Gamma_{2P}$ = 0.2~eV (labelled 3).
The fraction of $\Xi^-d$ atoms which decay is also plotted for case 1; cases 2
and 3 give almost indistinguishable values. The results are seen to exhibit
only small sensitivity to the choice of strong interaction parameters, but
considerable sensitivity to the value of $k_{STK}$. As the amount of Stark
mixing increases, the fraction of $S$ state capture increases, $P$ state capture
decreases and the number of $\Xi^-$ decays from the $\Xi^-d$ atom is reduced
as expected.

A particular feature of the Borie and 
Leon model is that it seems to over-estimate
 the amount of $P$ state capture. This problem has been discussed by
Batty \cite{Bat96} for $\bar{p}p$ atoms. He finds that the
introduction of an
additional normalization factor $K_0$, used for Stark transitions between $S$
and $P$ states only, {\it i.e.} $nS \rightarrow nP$ and $nP \rightarrow nS$
transitions, gives significantly improved fits to the X-ray yield data
with $K_0$ = 7.6 and much reduced values for the fraction of $P$ state
annihilation in liquid hydrogen.

\begin{figure}
\epsfig{file=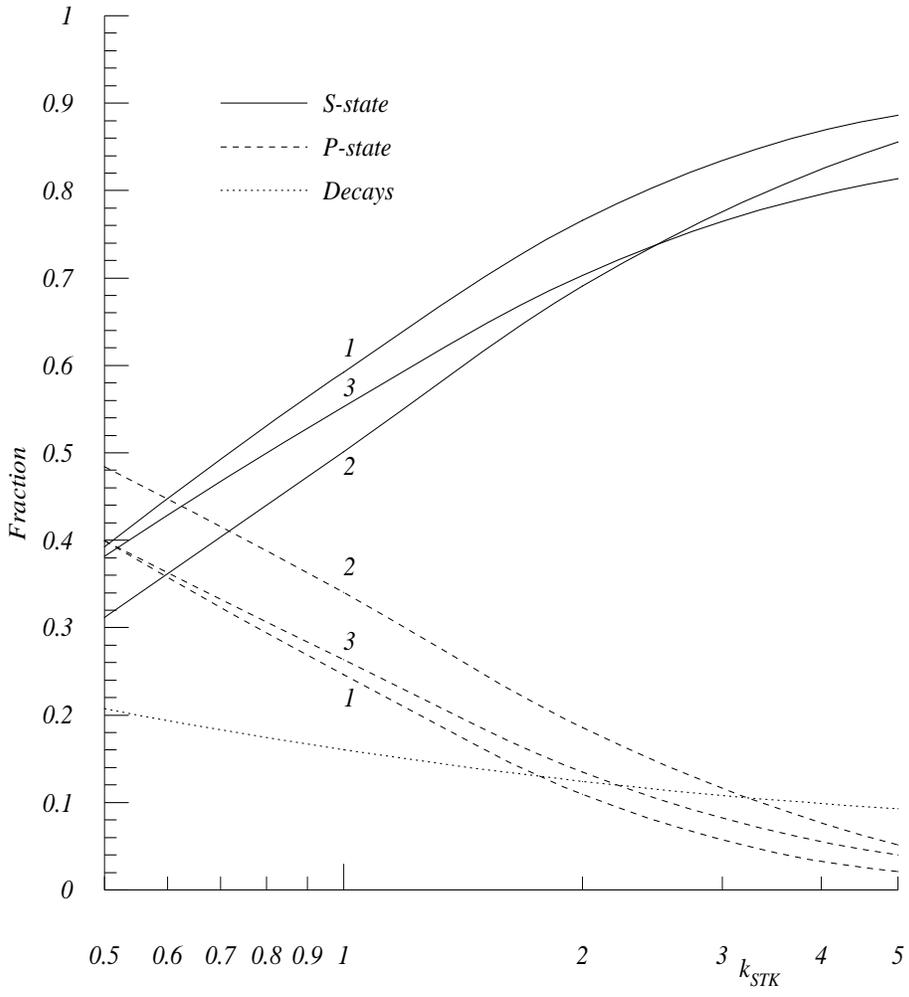,height=140mm,width=120mm,
bbllx=48,bblly=179,bburx=495,bbury=652}
\caption{Same as Fig. \protect{\ref {fig:Xid10}} but with the additional Stark
mixing parameter $K_0$ = 7.6.}
\label{fig:Xid76}
\end{figure}

Here we use the same model and again fix $K_0$ = 7.6. There is no
particular reason
for this choice of value for $K_0$ except that it gives a good fit to the
$\bar{p}p$ atom results. The resulting values for the
fraction of $S$ and
$P$ state capture and of $\Xi^-$ decays are plotted in Fig. \ref{fig:Xid76}
for the same values of the strong interaction parameters as used in Fig.
\ref{fig:Xid10} which is equivalent to the case with $K_0$ = 1.0.
Setting $K_0$ = 7.6 as in
Fig. \ref{fig:Xid76} decreases the $P$ state fraction and increases the $S$ state
fraction, as is to be expected for the increased amount of
Stark mixing between
$S$ and $P$ states and as was also shown by the $\bar{p}p$
calculations. The results
now show slightly more sensitivity to the strong interaction parameters
$\Gamma_{1S}$ and $\Gamma_{2P}$. However the fraction of $\Xi^-$
decays largely stays unaltered.

Using $S$ and $P$ state capture fractions
obtained in this way, with values for $\Gamma_{1S}$ and $\Gamma_{2P}$
given in Table II of Ref. \cite{ADo84}, the probability for
the production of the $H$ particle from $\Xi^-d$ atoms can be calculated
\cite{Low95}. Aerts and Dover \cite{ADo84} also give in Fig. 4
of their paper,
values, as a function of the mass of the $H$ particle $m_H$, for the
branching ratios $R_S$ and $R_P$ defined by,

\begin{equation}\label{eq:Hbrs}
R_S = \Gamma((\Xi^-d)_{1S} \rightarrow Hn)/\Gamma_{1S}
\end{equation}

and

\begin{equation}\label{eq:Hbrp}
R_P = \Gamma((\Xi^-d)_{2P} \rightarrow Hn)/\Gamma_{2P}.
\end{equation}

\begin{figure}
\epsfig{file=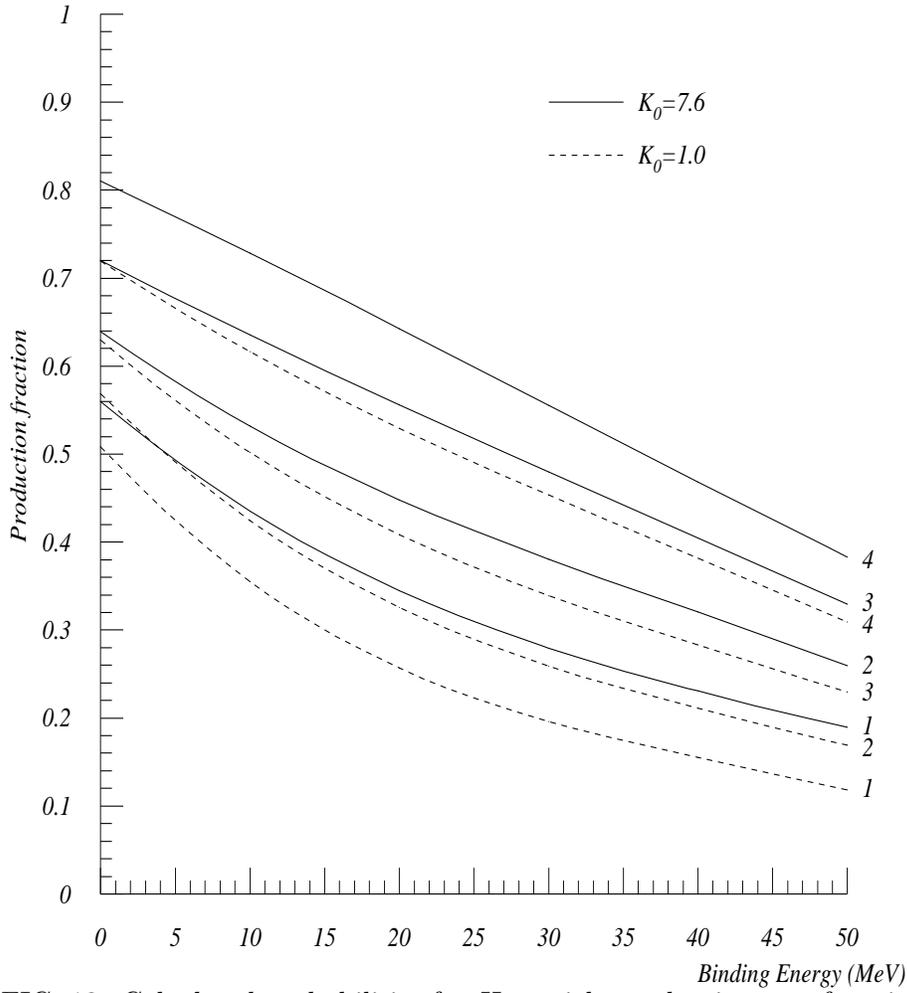,height=140mm,width=120mm,
bbllx=48,bblly=182,bburx=523,bbury=652}
\caption{Calculated probabilities for $H$ particle production as a function of
its binding energy $B(H)$. Curves labelled 1, 2, 3 and 4 correspond
to values of the Stark
mixing parameter $k_{STK}$ = 0.5, 1.0, 2.0 and 5.0 respectively. The dashed
and solid lines correspond to values of $K_0$ = 1.0 and 7.6 respectively.
(See text).}
\label{fig:Hprod}
\end{figure}

\noindent
Weighting the fractions of $S$ and $P$ state capture by
the corresponding $R_S$ and $R_P$ (values for model D in Ref.
\cite{ADo84} were used), then gives the total probability for $H$ particle
production as a function of
the binding energy $B(H)$. Here $B(H)$ is the binding energy of the $H$ particle
relative to the $\Lambda\Lambda$ mass. The results are shown in Fig.
\ref{fig:Hprod} for four different values of $k_{STK}$ = 0.5, 1.0, 2.0 and 5.0
(curves labelled 1, 2, 3, and 4 respectively) and for $K_0$ = 1.0 and 7.6
(dashed and solid lines respectively).

The results of \cite{ADo84} show that $H$ particle production is more likely
to proceed from $S$ states rather than $P$ states by a factor varying from 1.4 at
$m_H$ = 2.23~GeV ($B(H)$ = 0~MeV) to 14 at $m_H$ = 2.18~GeV
($B(H)$ = 50~MeV).
As a result the probability of $H$ particle production increases with Stark
mixing (increasing $k_{STK}$ or $K_0$) as can be clearly seen from Fig.
\ref{fig:Hprod}.

On the other hand the branching fractions
$R_S$ and $R_P$ both decrease as the binding energy $B(H)$ increases. However
the fraction of $S$ state capture increases somewhat from 0.46 to 0.57
and the $P$ state fraction decreases from 0.38 to 0.27 as $B(H)$ increases from 0
to 50~MeV. These
changes are largely due to the reduction in the $2P$ state strong interaction
width $\Gamma_{2P}$ from 0.61 to 0.25~eV as $B(H)$ increases;
the above capture fractions
were calculated for $k_{STK}$ = 1.0 and $K_0$ = 7.6. As a result,
due to the increase in $S$ state capture, the decrease
in $H$ particle production with increasing $B(H)$ is less rapid than might be
deduced solely from the overall reduction in the branching fractions for
$H$ particle production from the $\Xi^-d$ atom.

\section{Summary}
\label{sec:sum}

Almost no quantitative information on the interaction of $\Xi^-$
hyperons with nuclei is available at present and it is unlikely
that conventional measurements of particle energies
to investigate $\Xi^-$ hypernuclei will have
sufficient accuracy to alter this situation. In contrast, the usual
precision for measuring the energies of X-rays from transitions between
levels of hadronic atoms  offers the possibility of obtaining quantitative
information on the interaction of $\Xi^-$ hyperons with nuclei. In the
present work we have been guided by the successful observation and reasonably
precise measurement of strong interaction effects in $\Sigma^-$ atoms,
which has led to our best knowledge, so far, of the interaction of $\Sigma^-$
hyperons with nuclei.

Full atomic cascade calculations have been performed for $\Sigma^-$
and $\Xi^-$ atoms  and confirmed, as expected, that the processes within
these two hadronic atoms are very similar. The remaining major differences
are in the production reactions.
Whereas $\Sigma^-$ hyperons are produced by
the $p(K^-,\pi^+)\Sigma^-$ reaction at rest, the $p(K^-,K^+)\Xi^-$
reaction occurs at higher energies, thus causing decay losses during
the slowing down time
of the $\Xi^-$ particle to be non-negligible.
Prior to such an experiment it will be necessary to optimize
the experimental setup, which includes a hydrogen production target,
a heavy moderator such as Pb or W, the target to be studied and the
 detectors, both for X-rays and for the detection of the outgoing
$K^+$, which is essential in order to reduce background.

 In the present work we have confined ourselves to
studying the dependence of strong interaction shifts and widths
in $\Xi^-$ atoms on the
various parameters of the problem, including the nuclear charge.
Adopting an optical potential for the strong interaction between the $\Xi^-$
and the nucleus and assuming it to be attractive and 15-20 MeV deep with
an imaginary part of 1-3 MeV, we are able to propose four targets along
the periodic table, namely, F, Cl, I and Pb as suitable for X-ray measurements.
This also proves to be a sensible choice of targets for moderate changes
in $V_{opt}$. Nevertheless, even
 if the actual potential
turns out to be very different from the one used in the present calculations,
for example using potential G of Sec. \ref{sec:DD},
only relatively small changes will result,
because the whole phenomenon of hadronic atoms is dominated
by the Coulomb interaction.
Starting from a $G$-matrix based $\Xi N$ interaction, we constructed
a version of $V^{(\Xi)}$ (potential G) which would indeed lead to
a slightly different choice of optimal targets. It was shown that
the qualitative features of such a potential can be deduced by an
inversion procedure, where one fits a phenomenological DD potential
to the strong interaction data deduced from $\Xi^-$ atom experiments.
Although we do not consider potential G as a realistic one, this example
served to demonstrate the feasibility of the inversion procedure.

Finally, the special case of $\Xi^-d$ atoms has been discussed in great detail,
in view of its role in current experiments aimed at the $H$ dibaryon.
The role of Stark mixing, its effect on $S$ and
$P$ state capture of $\Xi^-$ by the deuteron, together with estimates
of the resulting probability for producing the $H$ dibaryon
have been considered in detail.

\vspace{15mm}

Some of the early calculations on $\Xi^-d$ atoms were carried out in
collaboration with G. T. A. Squier. The estimates of $H$ particle production
from $\Xi^-d$ atoms were made by J. Lowe.
This research was supported  by THE ISRAEL SCIENCE FOUNDATION
founded by The Academy of Sciences and Humanities.

\begin{table}
\caption{Calculated binding energies and strong interaction widths (in keV)
of $3D$  $\Xi^-$ atomic states.}
\label{tab:Blight}
\begin{tabular}{lccc}
 & $^{12}$C & $^{14}$N & $^{16}$O \\
Coulomb only & 126.36, $\Gamma$=0.000 & 174.71, $\Gamma$=0.000 &
 230.90, $\Gamma$=0.000 \\
potential 1 & 126.39, $\Gamma$=0.012 & 174.81, $\Gamma$=0.052 &
231.32, $\Gamma$=0.226 \\
potential 2 & 126.38, $\Gamma$=0.010 & 174.78, $\Gamma$=0.040 &
231.17, $\Gamma$=0.167 \\
\end{tabular}
\end{table}

\begin{table}
\caption{Predictions for likely targets for a $\Xi^-$ atoms experiment.
Calculations are based on a $t\rho$ potential with $b_0=0.25+i0.04$ fm.
E$_x$ is the transition energy, $Y$ is the upper level relative yield.}
\label{tab:preds}
\begin{tabular}{lccccc}
target & F & Cl & Sn & I & Pb \\  \hline
transition & $4F\rightarrow3D$ & $5G\rightarrow4F$&$8J\rightarrow7I$&
           $8J\rightarrow7I$ & $10L\rightarrow9K$ \\
E$_x$ (keV) & 131.29 & 223.55 & 420.25 & 474.71 & 558.47 \\
 $Y$ & 0.31 & 0.37 & 0.76 & 0.43 & 0.58 \\
shift (keV) & 1.56 & 1.84 & 0.67 & 2.79 & 1.73 \\
width (keV) & 0.99 & 1.14 & 0.43 & 2.21 & 1.26 \\  
\end{tabular}
\end{table}

\begin{table}
	\caption{Parameters for the $G$-matrix interaction
         Eq. (\protect{\ref{eq:gxin}})}
        \label{tab:gxin}
	\begin{tabular}{lccc}
		$\beta_{i}$ (fm) & 2.0 & 0.9 & 0.5  \\
		\hline
		$a_{i}$ (MeV)  & $-2.595$ & $-200.7$ & 821.4  \\
		$b_{i}$ (MeV~fm) & 0.2632 & 63.94 & $-124.0$  \\
	\end{tabular}
\end{table}

 \begin{table}
 	\caption{Calculated shifts and widths (in keV) due to
          potential G, Eqs. (\protect{\ref{eq:gxin}},\protect{\ref{eq:fold}})
and Table
          \protect\ref{tab:gxin}.}
        \label{tab:tgts}
 	\begin{tabular}{lcccccccc}
 		atom & O(3D) & F(3D) & S(4F) & Cl(4F) & Sn(7I) & I(7I) & W(9K) &
 		Pb(9K)  \\
 		\hline
 		shift & 3.343 & 14.23 & 7.511 & 20.53 & 8.725 & 40.39 & 3.191 &
 		22.90  \\
 		width & 0.924 & 7.619 & 2.523 & 11.68 & 3.393 & 40.29 & 0.846 &
 		16.63  \\
 	\end{tabular}
 \end{table}

\end{document}